\documentclass[prb,showpacs,twocolumn,superscriptaddress]{revtex4-1}
\usepackage{amssymb}
\usepackage{graphicx}
\usepackage{amsmath,amssymb}

\begin{document}

\title{Kondo screening of Andreev bound states in an N-QD-S system}
\author{Lin Li}
\affiliation{Department of physics, Southern University of science and technology of China, Shenzhen 518005, China}
\author{Zhan Cao}
\affiliation{Center of Interdisciplinary Studies and Key Laboratory for Magnetism and
Magnetic Materials of the Ministry of Education, Lanzhou University, Lanzhou 730000, China}
\author{Tie-Feng Fang}
\affiliation{Center of Interdisciplinary Studies and Key Laboratory for Magnetism and
Magnetic Materials of the Ministry of Education, Lanzhou University, Lanzhou 730000, China}
\author{Hong-Gang Luo}
\affiliation{Center of Interdisciplinary Studies and Key Laboratory for Magnetism and
Magnetic Materials of the Ministry of Education, Lanzhou University, Lanzhou 730000, China}
\affiliation{Beijing Computational Science Research Center, Beijing 100084, China}
\author{Wei-Qiang Chen}
\affiliation{Department of physics, Southern University of science and technology of China, Shenzhen 518005, China}

\begin{abstract}
Motivated by experimental observation of the Kondo-enhanced Andreev
transport [R. S. Deacon \textit{et al.}, PRB \textbf{81}, 121308(R) (2010)]
in a hybrid normal metal-quantum dot-superconductor (N-QD-S) device, we
theoretically study the Kondo effect in such a device and clarify the
different roles played by the normal and superconducting leads. Due to the
Andreev reflection in a QD-S system, a pair of Andreev energy levels form in
the superconducting gap, which is able to carry the magnetic moment if the
ground state of the QD is a magnetic doublet. In this sense, the Andreev
energy levels play a role of effective impurity levels. When the normal lead
is coupled to the QD-S system, on the one hand, the Andreev energy levels
broaden to form the so-called Andreev bound states (ABSs), on the other
hand, it can screen the magnetic moment of the ABSs. By tuning the couplings
between the QD and the normal (superconducting) leads, the ABSs can simulate
the Kondo, mixed-valence, and even empty orbit regimes of the usual
single-impurity Anderson model. The above picture is confirmed by the
Green's function calculation of the hybrid N-QD-S Anderson model and is also
able to explain qualitatively experimental phenomena observed by Deacon
\textit{et al.}. These results can further stimulate related experimental
study in the N-QD-S systems.
\end{abstract}

\maketitle

\section{Introduction}

The Kondo physics resulted from the screening of the local moment by
conduction electrons \cite{Kondo1964} provides an ideal platform to explore
the many-body correlations and their interactions. Below certain temperature
(the Kondo temperature $T_K$) the Kondo physics is characterized by a
resonance developed near the Fermi energy level. When the conduction
electrons consist of a conventional superconducting quasi-particles
described by a gapped excitation spectrum, the interplay between the
magnetism and superconductivity creates the so called Yu-Shiba-Rusinov (YSR)
bound state. \cite{Yu1965,Shiba1968,Rusinov1969,Flatte1997} On the other
hand, the competition between the Kondo and superconducting correlation also
determines the ground state of such systems. It is a magnetic doublet if the
superconducting correlation dominates over the Kondo effect, otherwise, the
ground state is a spin singlet state. \cite{Franke2011}

\begin{figure}[tbp]
\center{\includegraphics[clip=true,width=0.9\columnwidth]{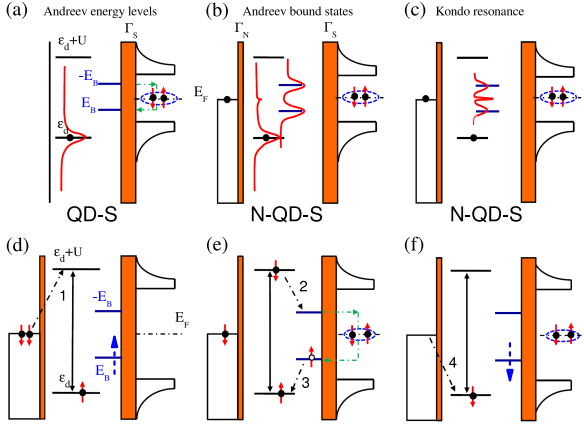}}
\caption{(a)-(c) The schematic ABSs and Kondo resonance in the N-QD-S
device. (a) In the absence of the normal lead($\Gamma_N = 0$), the ABSs
manifest as discrete Andreev energy levels located at $\pm E_B$. (b) Once
the normal lead is switched on ($\Gamma_N \neq 0$), the discrete Andreev
energy levels broaden to form the ABSs with the width in the order of $%
\Gamma_N$. (c) The Kondo resonance occurs once the temperature decreases to
certain characteristic temperature $T_K$. (d)-(f) The schematic picture of
the Kondo resonance lead by an effective spin-flip cotuneling process marked
by 1 to 4. (d) The initial state: a spin-up electron on the dot level $%
\protect\varepsilon_{d}$ forms a magnetic moment due to large $U$. The
strong coupling with the superconducting lead induces the ABSs, which are
able to carry this magnetic moment, as shown as blue dashed arrow. ``1"
denotes the virtual process of double occupancy of the dot. (e) The
intermediate states: by the process of ``2", a Cooper pair forms in the
superconducting lead and the spin-up electron combines with a hole
generated by Andreev reflection, denoted by the process ``3". (f) The final
state: one spin-down electron tunnels into the dot level denoted by the
process ``4", equivalently, the spin in the dot level is reversed.
Meanwhile, the effective magnetic moment in the ABSs is also reversed
effectively. As a result, the effective magnetic moment in the ABSs has been
screened by the electrons in the normal lead, which leads to the Kondo
resonance in the N-QD-S device.}
\label{fig1}
\end{figure}

In a junction consisting of metal and superconductor, the transport shows a
unique feature, namely, the Andreev reflection. When a low-energy electron
is incident on the interface, a Cooper pair is injected into the
superconductor and as a result, a hole with opposite spin is generated to go
back into the metal. In a confined geometry, this process is able to create
a pair of discrete Andreev energy levels [see Fig.\,\ref{fig1} (a)] in the
superconducting gap, as proposed theoretically. \cite%
{Beenakker1992,Zazunov2003,Avishai2003,Ossipov2007,Zhang2008,Skoldberg2008,Meng2009}
When a normal metal electrode is additionally coupled to the system [see
Fig.\,\ref{fig1} (b)], the discrete Andreev energy levels can broaden to
form the so-called Andreev bound states (ABSs). Recently, the individual
ABSs have been clearly identified in a semiconductor quantum dot (QD)
connected with normal and superconducting leads (N-QD-S) \cite%
{Deacon2010a,Chang2013,Lee2014} and in a superconductor-QD hybrid with
graphene. \cite{Dirks2011} In a successive paper, \cite{Deacon2010b} Deacon
\textit{et al.} further found that a Kondo resonance occurs around the Fermi
level if the coupling between the QD and the normal lead ($\Gamma_N$) is
much smaller than the coupling between the QD and the superconducting lead ($%
\Gamma_{S}$), as shown schematically in Fig.\,\ref{fig1} (c). However, in
the opposite case, namely, $\Gamma_{N}\gg\Gamma_{S}$, the Kondo resonance
near the Fermi level has not been observed. Furthermore, once the
superconducting gap is killed by applied magnetic field, the Kondo peaks
with Zeeman splitting have been observed. Though there do exist many
theoretical works to investigate the Kondo effect in the N-QD-S device, \cite%
{Clerk2000,Sun2001,Krawiec2004,Avishai2004,Domanski2008,Koerting2010,Yamada2011,Zitko2015,Domanski2015}
some fundamental issues involved in the Deacon \textit{et al.}'s experiment
can not be clearly explained. For example, if the superconducting lead is
available, one can not observe the Kondo resonance if $\Gamma_{N}\gg%
\Gamma_{S}$ and oppositely, the Kondo resonance occurs if $%
\Gamma_{S}\gg\Gamma_{N}$. When the superconductivity is killed by applied
magnetic field, the Kondo resonances with Zeeman splitting have been
observed in both cases. What is the underlying physics? More fundamentally,
what is the role played by the superconducting lead in such a hybrid N-QD-S
device?

In this work, we make an attempt to understand these experimental
observations. Fig.\,\ref{fig1} (d)-(f) represent our physical picture, which
shows schematically the microscopic mechanism of the Kondo resonance
observed in the experiment. \cite{Deacon2010b} The key point of this picture
is that the coupling between the QD and the superconducting lead induces a
pair of Andreev bound levels, which are able to carry effective magnetic
moment [denoted by blue dashed arrow in Fig.\,\ref{fig1} (d)-(f)] originated
from the single occupancy of the dot level if the ground state of the QD is
a magnetic doublet. In the spin-flip cotunneling process of the QD,
equivalently, the effective magnetic moment involved in the ABSs is
reversed. In this sense, the effective magnetic moment of the ABSs is
screened by the conduction electrons in the normal lead, which leads to the
Kondo resonance observed in the experiment. \cite{Deacon2010b} Differently
from the Kondo effect in the usual normal metal-quantum dot-normal metal
(N-QD-N) device, the spin-flip cotunneling process here is associated with
the Andreev reflection process, in which the effective magnetic moment
determined by the position of the ABSs plays an important role in such a
device. Therefore, one can image that the position of the ABSs can be tuned
by changing various parameters of the N-QD-S device, as a result, this
device can simulate various regimes involved in the single-impurity Anderson
model (SIAM),\cite{Anderson1961} for example, the Kondo regime, the
mixed-valence regime, and even the empty orbit regime, as discussed in
detail later. Our results can explain qualitatively the experimental
observations and thus clarify the questions mentioned above.

\section{Model and formalism}

To confirm above physical picture, in the following we consider the N-QD-S
device described by a hybrid Anderson impurity model. The Hamiltonian reads
\begin{equation}
H=\sum_{\beta = N,S} H_{\beta} + H_{d} + H_{V},  \label{Hamiltonian}
\end{equation}%
where $H_{d}=\sum_{\sigma }\varepsilon_{d\sigma}d_{\sigma
}^{\dagger}d_{\sigma}+U n_{d\uparrow} n_{d\downarrow}$ describes the QD with
the energy level $\varepsilon_{d\sigma}$ and the on-site repulsion
interaction $U$. $H_{\beta}=\sum_{k\sigma}\varepsilon_{k\beta}c_{k\sigma%
\beta }^{\dagger }c_{k\sigma\beta} + \delta_{\beta,S}\Delta
\sum_{k}(c_{k\uparrow\beta}^{\dagger}c_{-k\downarrow\beta}^{\dagger} + H.c.)$
is the Hamiltonian of the normal $(\beta = N)$ and the superconducting $%
(\beta = S)$ leads, which are contacted to the QD by the coupling term $%
H_{V}=\sum_{k\sigma\beta}\left(V_{k\beta}c_{k\sigma\beta}^{\dagger
}d_{\sigma }+H.c.\right)$. Here $\Delta$ is superconducting gap and $%
V_{k\beta}$ is the coupling strength with the $\beta$-lead.

Due to the superconducting proximity effect, it is convenient to express the
retarded dot Green's function (GF) by using the Nambu representation,
namely, $\hat{G}_{d\sigma }(t,t^{\prime })=\langle \langle \hat{\Psi
}_{\sigma }(t);\hat{\Psi}_{\sigma}^{\dagger }(t^{\prime })\rangle \rangle $
with $\hat{\Psi}_{\sigma }^{\dagger }=(d_{\sigma }^{\dagger },d_{\bar{\sigma%
}})$ and $\hat{\Psi} _{\sigma }=(\hat{\Psi}_{\sigma }^{\dagger })^{\dagger
} $. In the frequency space, the GF has a form of Dyson equation
\begin{equation}
\left[ \hat{G}_{d\sigma }\left( \omega \right) \right] ^{-1}=\left[ \hat{G}%
_{d\sigma }^{0}(\omega )\right] ^{-1}-\hat{\Sigma}_{\sigma }^{U}(\omega ),
\label{G}
\end{equation}%
where $\left[ \hat{G}_{d\sigma }^{0}(\omega )\right] ^{-1}=\hat{I}\omega -%
\hat{\sigma}_{z}\text{diag}(\varepsilon _{d\sigma },\varepsilon _{d\bar{%
\sigma}})-\hat{\Sigma}^{0}(\omega )$. Here $\hat{\Sigma}^{0}(\omega )$ is
the non-interacting self-energy with the components $\hat{\Sigma}%
_{11}^{0}(\omega )=$ $\hat{\Sigma}_{22}^{0}(\omega )=-i(\Gamma _{N}+\Gamma
_{S}\rho _{S}(\omega ))=-i\Gamma ({\omega })$, $\hat{\Sigma}_{12}^{0}(\omega
)=\hat{\Sigma}_{21}^{0}(\omega )=i\frac{\sigma \Delta }{\omega }\Gamma _{S}\rho _{S}({\omega }),$
$\rho _{S}\left( \omega \right) =\frac{\left\vert
\omega \right\vert \theta \left( \left\vert \omega \right\vert -\Delta
\right) }{\sqrt{\omega ^{2}-\Delta ^{2}}}+\frac{\omega \theta \left( \Delta
-\omega \right) }{i\sqrt{\Delta ^{2}-\omega ^{2}}},$ and the coupling $%
\Gamma _{\beta }=\pi |V_{k\beta }|^{2}/(2D)$.\cite{Sun2000,Tanaka2007} Here
we take the coupling matrix element $V_{k\beta }$ to be $k$-independent and
shall restrict to the limit $|\varepsilon _{k\beta }|\ll D$, where $D$ is
the half-bandwidth. The self-energy $\hat{\Sigma}_{\sigma }^{U}(\omega )$ is
due to the on-site repulsion interaction, denoting correlation effect.\cite{Domanski2008b,Baranski2011,Vecino2003} Following the notations introduced in Refs. [
\onlinecite{Bauer2007}] and [\onlinecite{Bulla1998}], the self-energy $\hat{\Sigma}_{\sigma }^{U}(\omega )$ has a matrix form $\hat{\Sigma}_{\sigma
}^{U}(\omega )=U\hat{F}_{d\sigma }(\omega )[\hat{G}_{d\sigma }\left( \omega
\right) ]^{-1}$, where
\begin{equation}
\hat{F}_{d\sigma }\left( \omega \right) =\left(
\begin{array}{cc}
\langle \langle d_{\sigma }n_{d\bar{\sigma}};d_{\sigma }^{\dagger }\rangle
\rangle & \langle \langle d_{\sigma }n_{d\bar{\sigma}};d_{\bar{\sigma}%
}\rangle \rangle \\
-\langle \langle d_{\bar{\sigma}}^{\dagger }n_{d\sigma };d_{\sigma
}^{\dagger }\rangle \rangle & -\langle \langle d_{\bar{\sigma}}^{\dagger
}n_{d\sigma };d_{\bar{\sigma}}\rangle \rangle%
\end{array}%
\right).  \label{FG}
\end{equation}%
In the framework of equation of motion of GF, this self-energy cannot be
obtained exactly, and one has to employ truncation approximation. As shown by
Eq.(\ref{CE}) in Appendix, the high-order GFs in Eq. (\ref{FG}) can be treated conventionally by a cluster expansion, \cite{Luo1999}
where the connected GFs like $\langle \langle d_{\sigma }n_{d\bar{\sigma}};d_{\sigma }^{\dagger }\rangle\rangle _{c}$ contains the contribution from the more higher order correlation effect, which can be truncated at different approximate level. As truncated in the first order, the Hartree-Fock approximation (HFA) can be reached by
neglecting the connected GFs $\hat{F}_{d\sigma }(\omega )_c$ as defined in Appendix. In this case the
self-energy $\hat{\Sigma}_{\sigma }^{U}(\omega )$ is approximated by
\begin{equation}
\hat{\Sigma}_{\sigma }^{U}(\omega )\approx \hat{\Sigma}_{\sigma
}^{HF}(\omega )=U\left(
\begin{array}{cc}
\langle n_{d\bar{\sigma}}\rangle & \langle d_{\bar{\sigma}}d_{\sigma }\rangle
\\
\langle d_{\sigma }^{\dagger }d_{\bar{\sigma}}^{\dagger }\rangle & -\langle
n_{d\sigma }\rangle%
\end{array}%
\right) ,  \label{GF-HF}
\end{equation}%
where the occupation $\langle n_{d\bar{\sigma}}\rangle =-\frac{1}{\pi }\int
f\left( \omega \right) \text{Im}[\hat{G}_{d\bar{\sigma}}\left( \omega
\right) ]_{11}d\omega $, $f\left( \omega \right) $ is the Fermi distribution
function.\cite{Yamada2011,Vecino2003,Osawa2008} $\langle d_{\sigma
}^{\dagger }d_{\bar{\sigma}}^{\dagger }\rangle $ is the pairing correlation
function of the QD, which can be explicitly evaluated with $\langle
d_{\sigma }^{\dagger }d_{\bar{\sigma}}^{\dagger }\rangle =-\frac{1}{\pi }%
\int f\left( \omega \right) \text{Im}[\hat{G}_{d\sigma }\left( \omega
\right) ]_{21}d\omega$, where the anomalous GF obtained is
\begin{equation}
\left[ \hat{G}%
_{d\sigma }\left( \omega \right) \right] _{21}=\frac{\hat{\Sigma}_{21}^{0}\left( \omega \right)+U\langle d_{\sigma }^{\dagger }d_{%
\bar{\sigma}}^{\dagger }\rangle }{\omega+\varepsilon _{d\bar{\sigma}}+i\Gamma(\omega) +U\langle n_{d\sigma }\rangle }\left[ \hat{G}%
_{d\sigma }\left( \omega \right) \right] _{11}. \label{G12R}
\end{equation}
[see Eq.(\ref{AGF-GF}) in Appendix]
The HFA has been used to treat the level-crossing
quantum phase transition between the BCS-singlet and the magnetic doublet
states, \cite{Lee2014,Vecino2003,Cuevas2001, Benjamin2007} and it is also
useful to determine the magnetic regime of the N-QD-S device as following
the scheme of Anderson. \cite{Anderson1961} However, it is well-known that
in the HFA level the Kondo effect is absent, and thus one needs to consider
the high-order GF's (see, e.g., Refs.[\onlinecite{Luo1999}] and [\onlinecite{Lacroix1981}]).

To capture the Kondo physics, one only needs to consider the contribution of diagonal
connected GFs, such as $\langle \langle d_{\sigma }n_{d\bar{\sigma}%
};d_{\sigma }^{\dagger }\rangle \rangle _{c}$, \cite{Domanski2008b,Baranski2011,Baranski2013} because the non-diagonal conponents
like $\langle \langle d_{\bar{\sigma}}^{\dagger }n_{d\sigma };d_{\sigma
}^{\dagger }\rangle \rangle _{c}$ will introduce additional influence of the correlation
effect on the superconducting lead, in which we are not interested. Here we take the Lacroix's truncation approximation,\cite{Lacroix1981}
as widely used in the literature. \cite{Shiau2007,Qi2008,Feng2009,Fang2010,Lim2013,Krychowski2014,Li2014} For
details, one can refer to Appendix. Under this approximation, the ``11"-component of the dot GF reads
\begin{equation}
\left[ \hat{G}%
_{d\sigma }\left( \omega \right) \right] _{11}
=\frac{1}{R_{\sigma }(\omega )-U[Q_{\sigma }(\omega )+T_{\sigma }(\omega )]/P_{\sigma }(\omega )},  \label{GD11}
\end{equation}
where the following notations have been introduced for clarity,
\begin{widetext}
\begin{eqnarray}
R_{\sigma }(\omega ) &=&\omega -\varepsilon _{d\sigma }+i\Gamma (\omega
)-U\langle n_{d\bar{\sigma}}\rangle -\frac{( \hat{\Sigma}%
_{21}^{0}\left( \omega \right) +U\langle d_{\bar{\sigma}}d_{\sigma }\rangle
) ( \hat{\Sigma}_{21}^{0}\left( \omega \right) +U\langle
d_{\sigma }^{\dagger }d_{\bar{\sigma}}^{\dagger }\rangle) }{\omega
+\varepsilon _{d\bar{\sigma}}+i\Gamma (\omega )+U\langle n_{d\sigma }\rangle
},  \label{QS} \\
P_{\sigma }(\omega ) &=&\omega -\varepsilon _{d\sigma }-U\left( 1-\langle
n_{d\bar{\sigma}}\rangle \right) +3i\Gamma (\omega )+U(A_{1\sigma}(\omega)-A_{2\sigma}(\omega)),  \label{PS} \\
Q_{\sigma }(\omega ) &=&\left( \omega -\varepsilon _{d\sigma }-U\langle n_{d\bar{\sigma}}\rangle
\right) (A_{1\sigma}(\omega)-A_{2\sigma}(\omega))+U[\langle n_{d\bar{\sigma}}\rangle ( 1-\langle
n_{d\bar{\sigma}}\rangle)-\langle d_{\bar{\sigma}%
}d_{\sigma}\rangle \langle d_{\sigma}^{\dag}d_{\bar{\sigma}%
}^{\dagger}\rangle] \notag \\
&&-2i\Gamma (\omega )\langle n_{d\bar{\sigma}%
}\rangle - (B_{1\sigma}(\omega)+B_{2\sigma}(\omega)),  \label{RS} \\
T_{\sigma}\left(  \omega\right)&=&\left[  \varepsilon_{d\sigma
}+\varepsilon_{d\bar{\sigma}}+U\left(  1+\left\langle n_{d\sigma}\right\rangle
-\left\langle n_{d\bar{\sigma}}\right\rangle \right)  \right]  \frac{ \left\langle
d_{\bar{\sigma}}d_{\sigma}\right\rangle(\hat{\Sigma}_{21}^{0}\left( \omega \right)
+U\langle d_{\sigma}^{\dagger}d_{\bar{\sigma}}^{\dag}\rangle)}{\omega
+\varepsilon _{d\bar{\sigma}}+i\Gamma (\omega )+U\left\langle n_{d\sigma
}\right\rangle },\label{TS}
\end{eqnarray}
with
\begin{equation}
A_{i\sigma }\left( \omega \right) =\frac{i}{2\pi ^{2}}\sum_{\beta \left(
=N,S\right) }\int \int d\omega ^{\prime
}d\varepsilon \Gamma _{\beta }f\left( \omega ^{\prime }\right) \frac{\frac{\left( z_{+}^{\prime
}+\varepsilon \right) \left[ \hat{G}_{d\bar{\sigma}}\left( \omega ^{\prime
}\right) \right] _{11}-\delta _{\beta ,S}\bar{\sigma}\Delta \left[ \hat{G}_{d%
\bar{\sigma}}\left( \omega ^{\prime }\right) \right] _{21}}{\left(
z_{+}^{\prime }-\varepsilon \right) \left( z_{+}^{\prime }+\varepsilon
\right) -\delta _{\beta ,S}\Delta ^{2}}-\frac{\left( z_{-}^{\prime
}+\varepsilon \right) \left[ \hat{G}_{d\bar{\sigma}}\left( \omega ^{\prime
}\right) \right] _{11}^{\ast }-\delta _{\beta ,S}\bar{\sigma}\Delta \left[
\hat{G}_{d\bar{\sigma}}\left( \omega ^{\prime }\right) \right] _{21}^{\ast }%
}{\left( z_{-}^{\prime }-\varepsilon \right) \left( z_{-}^{\prime
}+\varepsilon \right) -\delta _{\beta ,S}\Delta ^{2}}}{z_{+}-\varepsilon
_{i\sigma }},  \label{AIS}
\end{equation}
and
\begin{equation}
B_{i\sigma }\left( \omega \right) \approx \frac{1}{\pi }\int d\varepsilon \frac{\Gamma _{N}f\left(
\varepsilon \right) [ 1-i\Gamma _{N}( \hat{G}_{d\bar{\sigma}%
}\left( \varepsilon \right) ) _{11}] }{z_{+}-\varepsilon
_{i\sigma }}+\frac{i}{2\pi ^{2}}\int
\int  d\omega ^{\prime }d\varepsilon \Gamma _{S}\left( \varepsilon \right)f\left( \omega ^{\prime }\right)
\frac{\frac{z_{+}^{\prime }+\varepsilon }{\left( z_{+}^{\prime }-\varepsilon
\right) \left( z_{+}^{\prime }+\varepsilon \right) -\Delta ^{2}}-\frac{%
z_{-}^{\prime }+\varepsilon }{\left( z_{-}^{\prime }-\varepsilon \right)
\left( z_{-}^{\prime }+\varepsilon \right) -\Delta ^{2}}}{z_{+}-\varepsilon
_{i\sigma }},  \label{BIS}
\end{equation}
\end{widetext}
where $z_{\pm}=\omega \pm i\eta$ ($\eta\rightarrow 0^{+}$), and $\varepsilon _{1\sigma }=\varepsilon +\varepsilon _{d\sigma }-\varepsilon _{d%
\bar{\sigma}}$, $\varepsilon _{2\sigma }=-\varepsilon +\varepsilon _{d%
\bar{\sigma}}+\varepsilon _{d\sigma }+U$.
The GF's in Eqs.(\ref{G12R}) and (\ref{GD11}) are closed and can be calculated self-consistently.
The equation of motion treatment is simple enough, but it is
quite good at high temperature, and it can capture qualitatively the Kondo
physics even at low temperatures. The main goal of the present work is to
explain not quantitatively but qualitatively the experimental observations
given by Deacon \emph{et al.}, \cite{Deacon2010b} and thus the result of the
equation of motion is sufficient to confirm our physical picture.

\section{Numerical results}

In order to obtain the Kondo resonance, we discuss the local density of
states $\rho_{d}(\omega)=-\frac{1}\pi \sum_{\sigma}\text{Im}[\hat{G}
_{d\sigma}(\omega)]_{11}$ of the QD. In our calculations, the
superconducting gap $\Delta$ is taken as the units of energy, and the half
bandwidth $D=20\Delta$.

\begin{figure}[tbp]
\center{\includegraphics[clip=true,width=0.9\columnwidth]{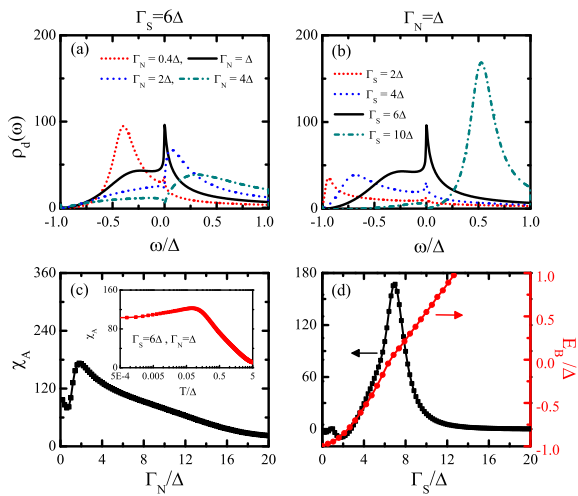}}
\caption{(a)-(b) The evolution of Kondo resonance and ABS in the local
density of states by tuning the couplings of $\Gamma_{N}$ and $\Gamma_{S}$,
respectively. (c)-(d) The spin susceptibility of the ABS as a function of
the couplings of $\Gamma_{N}$ and $\Gamma_{S}$. $\Gamma_{S}=6\Delta$ in (a)
and (c) and $\Gamma_{N}=\Delta$ in (b) and (d). The red dotted-line in (d)
indicates the coupling ($\Gamma_{S}$) dependent of the Andreev bound level $
E_{B}$. Other parameters used are the dot level $\protect\varepsilon_{d
\protect\sigma}=-5\Delta$, the Coulomb interaction $U=40\Delta$, and the
temperature $T=0$. The inset of (c) is the temperature dependent
susceptibility.}
\label{fig2}
\end{figure}

To extract the role of normal and superconducting leads in the Kondo
screening, we show the development and evolution of Kondo resonance by
tuning the coupling $\Gamma_{N}$ and $\Gamma_{S}$ in Fig.\,\ref{fig2} (a)
and (b), respectively. The position of ABSs level $E_{B}$, corresponding to
the pair breaking excitation due to the local spin, is mainly determined by
the coupling $\Gamma_{S}$. When the level $E_{B}$ is situated below the
Fermi level, the ground state of QD is a magnetic doublet state. \cite%
{Franke2011,Li2014,Sellier2005} Once the normal lead is applied, the ABSs
are broadened and at the same time, a Kondo resonance at the Fermi level is
developed, as shown in Fig.\,\ref{fig2} (a). For fixed $\Gamma_{S}=6\Delta$,
the Kondo resonance evolves from a peak ($\Gamma_{N}=0.4\Delta, \Delta$)
into an asymmetric structure ($\Gamma_{N}=2\Delta, 4\Delta$) by increasing
the coupling $\Gamma_{N}$. The asymmetric structures originate from the Fano
resonance created by the interference of Kondo resonance and significant
broadened ABSs.\cite{Li2015} These are the typical behaviors of the SIAM
ranging from the Kondo to mixed-valence regime.\cite{Costi1994,Luo2004} The
above results indicate that the ABSs can be recognized as a localized level
with an effective moment. Physically, the moment of ABSs originates from the
quasi-particle excited by the pair-broken scattering of local spin. The
level $E_{B}$ can be broadened and screened by applied normal lead,
accordingly, creating a significant pronounced Kondo resonance peak at the
Fermi level. In Fig.\,\ref{fig1}, we showed the microscopic mechanism of
Kondo screening in the N-QD-S device, where the screening of the local spin
in the QD can be equivalently seen as a process that the effective moment of
ABSs is screened through the associated Andreev reflection.

In order to clarify the magnetic nature of the ABSs and its screening
process, in Fig.\,\ref{fig2} (c), we show the spin susceptibility of the ABS
defined by $\chi_{A}=\frac{g\mu_{B}(n_{A\uparrow}-n_{A\downarrow})}{h}%
\vert_{h\rightarrow0}$, where $n_{A\sigma}=-\frac{1}{\pi}\int_{-\Delta
}^{+\Delta } f(\omega)\text{Im}[\hat{G}_{d\sigma}(\omega)]_{11}d\omega$ is
the occupation of ABS existing in the superconducting gap. In addition, $h$
is a weak applied magnetic filed, $g$ is the Land\'e factor, $\mu_{B}$ is
the Bohr magneton, and we take $g\mu_{B}=1$. \textbf{It is seen that the susceptibility 
can be suppressed by increasing $\Gamma_{N}$, which indicates the Kondo screening 
of the effective moment of ABS by the conduction electrons in normal lead.} 
\textbf{While the increasing in the susceptibility with $\Gamma_{N} \lesssim 2\Delta$ may 
correspond to the enhancement of local moment in the ABS.} 
In addition, the inset of Fig.\,\ref{fig2} (c) shows the temperature dependence
of susceptibility $\chi_{A}$, which is qualitatively consistent with
the general Kondo screening behavior of local moment. \cite{Fang2015}
Therefore, the Kondo effect in the N-QD-S can be simply described by the
Kondo screening of the ABSs.

It is well-known that changing the Anderson impurity system from the Kondo
regime, to the mixed-valence regime, and even to the empty orbit regime can
be realized by tuning the impurity level. Here we fix $\Gamma_N=\Delta$ and
tune the level position of $E_{B}$ by increasing $\Gamma_{S}$ from $2\Delta$
to $10\Delta$, as shown in Fig.\,\ref{fig2} (b). It is seen that the Kondo
resonance takes place when the ground state of QD is a magnetic doublet
state ($\Gamma_{S}=2\Delta,4\Delta,6\Delta$). In contrast, the Kondo
resonance disappears as the ABSs shift above Fermi level (in empty orbit
regime), because the ground state is a Kondo singlet state ($\Gamma_{S}=10\Delta$)
even without the normal lead.
Likewise, we calculate $\chi_A$ as a
function of $\Gamma_S$, as shown in Fig.\,\ref{fig2} (d) for the
squared-line.
With increasing $\Gamma_S$, $\chi_A$ increases firstly, and
then decreases dramatically. The former can be understood as follows. When $%
\Gamma_S$ is switched on, the ABSs begins to form at the gap edge but with
small weight, and the effective local moment of ABSs gradually develops (the negative $\chi_A$ for some small $\Gamma_S$ may be artificial of the truncation approximation).
With increasing $\Gamma_S$, the ABSs begin to move toward the
Fermi level, as shown in Fig.\,\ref{fig2} (d) for dotted-line curve, and its
weight and as a result the effective magnetic moment, also increases. At the
same time, the effective magnetic moment would be screened by the conduction
electrons in the normal metal, leading to the Kondo resonance peak around the Fermi level, as shown in Figs. \ref{fig2}(a,b). Once the ABSs crosses the Fermi level, it seems
that the system enters into a mixed valence or an empty orbit regime, $\chi_A$ decreases
rapidly up to zero. In this sense, the coupling $\Gamma_S$ in the N-QD-S
device just plays a role of gate-voltage in the usual N-QD-N device.

\begin{figure}[tbp]
\includegraphics[clip=true,width=0.9\columnwidth]{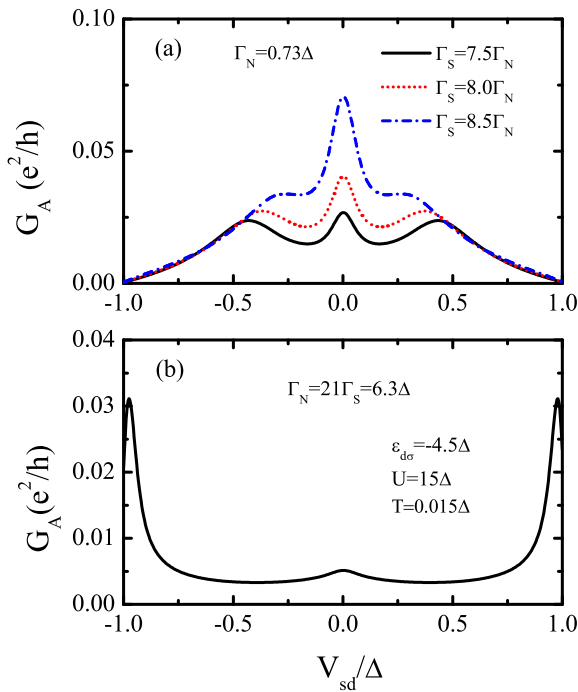}
\caption{The Andreev transport of the N-QD-S device for (a) $\Gamma_S =(7.5,
8.0, 8.5)\Gamma_N$ and (b) $\Gamma_N = 21\Gamma_S$. Other parameters used are
also similar to those in the Deacon \emph{et al.}'s experiment \protect\cite%
{Deacon2010b} with the dot level $\protect\varepsilon_{d\protect\sigma
}=-4.5\Delta$, the temperature $T=0.015\Delta$, and the Coulomb interaction $%
U=15\Delta$.}
\label{fig3}
\end{figure}

To confirm the above picture, we consider the Andreev transport in the
N-QD-S device. The Andreev conductance is $G_{A}=\frac{dI_{A}}{dV_{sd}}$
with $I_{A}=\frac{2e}{h}\int \Gamma^{2}_{N}\vert[\hat{G}_{d\sigma }\left(
\omega \right) ]_{21}\vert ^{2} [ f_{N}( \varepsilon -eV_{sd})
-f_{N}(\varepsilon+eV_{sd}) ]d\varepsilon$, as measured in the Deacon \emph{%
et al.}'s experiment.\cite{Deacon2010b} Fig.\,\ref{fig3} (a) and (b) show
the results for the cases $\Gamma_S \gg \Gamma_N$ and $\Gamma_N \gg \Gamma_S$%
, respectively. For the former one there exists three peaks, and the central
one is the Kondo resonance, which is qualitatively consistent with the
experimental observation. For the latter one the Kondo resonance is strongly
suppressed and the ABSs are located around the edge of the superconducting
gap, which is also similar to that observed in the experiment. Our results
give a physically reasonable explanation of the Deacon \emph{et al.}'s
experiment. \cite{Deacon2010b} The underlying physical mechanism is the
Kondo screening of the ABSs, which shows the constructive effect of the
superconducting lead on the Kondo transport. In order to further confirm
this, we compare three different couplings $\Gamma_S =(7.5, 8.0, 8.5)\Gamma_N$
in Fig. \ref{fig3}(a). It is found that the Kondo resonance is significantly
enhanced with the increasing coupling $\Gamma_S$, which is qualitatively
consistent with Domanski \emph{et al.}'s numerical renormalization group
calculation. \cite{Domanski2015}

\begin{figure}[tbp]
\includegraphics[clip=true,width=0.9\columnwidth]{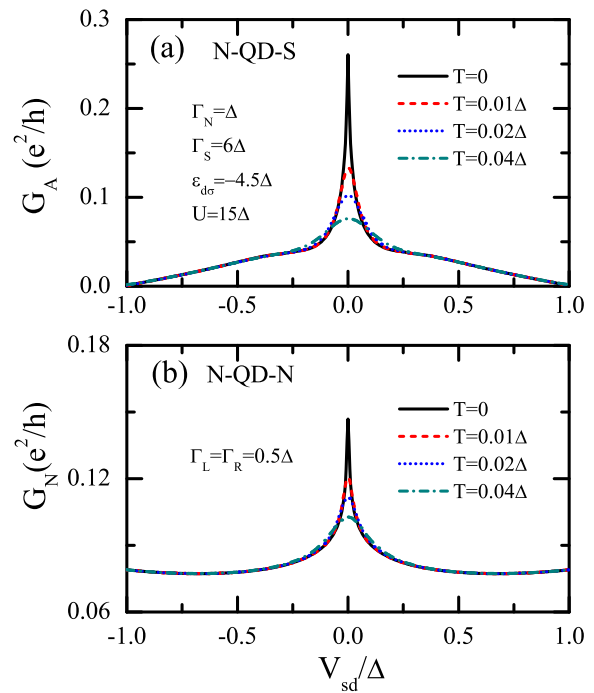}
\caption{(a) The temperature dependence of Kondo resonance peak in the
Andreev transport of the N-QD-S device. The results are obtained by using
the couplings of $\Gamma_{N}=\Delta$ and $\Gamma_{S}=6\Delta$. (b) The
temperature dependence of Kondo resonance in a general N-QD-N device by
using $\Gamma_{L}=\Gamma_{R}=0.5\Delta$. The other parameters used are $%
\protect\varepsilon_{d\protect\sigma}=-4.5\Delta$, and $U=15\Delta$.}
\label{fig4}
\end{figure}

The constructive role of the superconducting lead on the subgap Kondo transport can
also directly checked by comparing the general features of the Kondo
transport ($G_{A}$) for the N-QD-S and ($G_{N}$) for an N-QD-N devices, as
shown in Fig. \ref{fig4}(a) and (b), respectively. For the sake of
comparison, we take $\Gamma_L =\Gamma_N =0.5\Delta$. On one hand,
the conductance in Fig.\,\ref{fig4} (a) is greatly larger than that in Fig.\,\ref{fig4} (b),
suggesting the constructive effect of the superconducting lead on the Kondo transport,
which is in agreement with the experimental observation.
\cite{Deacon2010b} On the other hand, although in the both cases the Kondo
resonances are obviously underestimated due to the equation of motion method
we used, the Kondo peaks in both cases can be suppressed dramatically by
increasing temperature. However, up to $T = 0.04\Delta$, in the N-QD-N case,
a broad peak around the zero bias is still visible, and on the contrary, in
the N-QD-S case, the Kondo peak is almost completely killed, which indicates
that the Kondo temperature in the N-QD-S should be smaller than that of the
N-QD-N. In this sense, our result is also qualitatively compatible with the
experimental observation of $T^{S}_{K}\ll T^{N}_{K}$. \cite{Deacon2010b}

\section{Conclusion}

In conclusion, we studied the nature of the Kondo screening in the N-QD-S
device. Both the normal and the superconducting leads can contribute to the
Kondo effect, but they play different roles in the Kondo screening of local
spin. The coupling between the local spin and superconductor induces a pair
of Andreev energy levels in the superconducting gap, and it does not
directly contribute to the Kondo screening, as the Kondo resonance peak can
not be directly observed in the S-QD-S system. \cite{Franceschi2010,Maurand2012,Pillet2013,Kim2013,Kumar2014} The ABSs possess
effective moment when the ground state of the QD is a magnetic doublet
state. When the normal lead is switched on, the Andreev energy levels
broaden to form ABSs and at the same time, a Kondo screening happens due to
the normal leads. Thus the Kondo resonance in the N-QD-S device can be
recognized as the screening of ABSs by the conduction electrons in normal
lead. The physical picture can explain qualitatively the experimental
observations and may further stimulate the related experimental studies.

\section*{Acknowledgement}

Lin Li acknowledges useful discussion with Hua Chen and Fu-Chun Zhang. This
work is supported by NSFC (Grants Nos. 11547110, 11325417, 11204186,
11274269, 11674139) of China, and Natural Science
Foundation of Guangdong Province of China 2014A030310137.

\section*{Appendix}
In this appendix, we present some main steps to show the truncation
approximations in obtaining the Green's functions (GFs) in the main text in the framework of
Hartree-Fock (HF) and Lacroix's approximations. The basic starting point is
the following equation of motion of the retarded Green's function \cite{Zubarev1960}
\begin{widetext}
\begin{equation}
\omega \langle \langle A;B\rangle \rangle =\langle \lbrack A,B]_{+}\rangle
+\langle \langle \lbrack A,H]_{-};B\rangle \rangle ,  \label{EM}
\end{equation}%
where the subscript $\pm$ stands for the anti-commutation (commutation)
relationship, and\ $\langle \langle A;B\rangle \rangle $ denotes the
retarded GF composed by the operators $A$ and $B$. From the Hamiltonian Eq.(%
\ref{Hamiltonian}) in the main text, one can obtain the GF's of the QD
\begin{eqnarray}
\omega \langle \langle d_{\sigma };B\rangle \rangle &=&\langle \lbrack
d_{\sigma };B]_{+}\rangle +\varepsilon _{d\sigma }\langle \langle d_{\sigma
};B\rangle \rangle +U\langle \langle d_{\sigma }n_{d\bar{\sigma}};B\rangle
\rangle +\sum_{k\beta }V_{k\beta }^{\ast }\langle \langle c_{k\sigma \beta
};B\rangle \rangle ,  \label{dG} \\
\omega \langle \langle c_{k\sigma \beta };B\rangle \rangle &=&\langle
\lbrack c_{k\sigma \beta };B]_{+}\rangle +\varepsilon _{k\beta }\langle
\langle c_{k\sigma \beta };B\rangle \rangle +V_{k\beta }\langle \langle
d_{\sigma };B\rangle \rangle +\sigma \Delta \delta _{\beta ,S}\langle
\langle c_{-k\bar{\sigma}\beta }^{\dagger };B\rangle \rangle ,  \label{cG} \\
\omega \langle \langle c_{-k\bar{\sigma}\beta }^{\dagger };B\rangle \rangle
&=&\langle \lbrack c_{-k\bar{\sigma}\beta }^{\dagger };B]_{+}\rangle
-\varepsilon _{-k\beta }\langle \langle c_{-k\bar{\sigma}\beta }^{\dagger
};B\rangle \rangle -V_{-k\beta }^{\ast }\langle \langle d_{\bar{\sigma}%
}^{\dagger };B\rangle \rangle -\bar{\sigma}\Delta ^{\ast }\delta _{\beta
,S}\langle \langle c_{k\sigma \beta };B\rangle \rangle ,  \label{cdG} \\
\omega \langle \langle d_{\bar{\sigma}}^{\dagger };B\rangle \rangle
&=&\langle \lbrack d_{\bar{\sigma}}^{\dagger };B]_{+}\rangle -\varepsilon _{d%
\bar{\sigma}}\langle \langle d_{\bar{\sigma}}^{\dagger };B\rangle \rangle
-U\langle \langle d_{\bar{\sigma}}^{\dagger }n_{d\sigma };B\rangle \rangle
-\sum_{k\beta }V_{-k\beta }\langle \langle c_{-k\bar{\sigma}\beta }^{\dagger
};B\rangle \rangle ,  \label{ddG}
\end{eqnarray}%
where $\sigma (\bar{\sigma})$ in the subscript represents the spin
orientation $\uparrow (\downarrow )$ or $\downarrow (\uparrow )$, while
those appear in the coefficients are set to be $\pm 1$ for $\uparrow
(\downarrow )$.

As choosing $B=d_{\sigma }^{\dagger }$ and removing the GFs of $\langle
\langle c_{k\sigma \beta };B\rangle \rangle $ and $\langle \langle c_{-k\bar{%
\sigma}\beta }^{\dagger };B\rangle \rangle $ in Eqs.(\ref{dG})-(\ref{ddG}),
one can obtain%
\begin{eqnarray}
\left( \omega -\varepsilon _{d\sigma }-\Gamma_1(\omega)\right) \langle \langle d_{\sigma
};d_{\sigma }^{\dagger }\rangle \rangle +\Delta _{1}\left( \omega \right)
\langle \langle d_{\bar{\sigma}}^{\dagger };d_{\sigma }^{\dagger }\rangle
\rangle  &=&1+U\langle \langle d_{\sigma }n_{d\bar{\sigma}};d_{\sigma
}^{\dagger }\rangle \rangle ,  \label{GE-1} \\
\left( \omega +\varepsilon _{d\bar{\sigma}}-\Gamma _{2}\left( \omega \right)
\right) \langle \langle d_{\bar{\sigma}}^{\dagger };d_{\sigma }^{\dagger
}\rangle \rangle -\Delta _{2}\left( \omega \right) \langle \langle d_{\sigma
};d_{\sigma }^{\dagger }\rangle \rangle  &=&-U\langle \langle d_{\bar{\sigma}%
}^{\dagger }n_{d\sigma };d_{\sigma }^{\dagger }\rangle \rangle .
\label{GE-2}
\end{eqnarray}%
Similarly, if one chooses $B=d_{\bar{\sigma}},$ we have%
\begin{eqnarray}
\left( \omega -\varepsilon _{d\sigma }-\Gamma _{1}\left( \omega \right)
\right) \langle \langle d_{\sigma };d_{\bar{\sigma}}\rangle \rangle +\Delta
_{1}\left( \omega \right) \langle \langle d_{\bar{\sigma}}^{\dagger };d_{%
\bar{\sigma}}\rangle \rangle  &=&U\langle \langle d_{\sigma }n_{d\bar{\sigma}%
};d_{\bar{\sigma}}\rangle \rangle ,  \label{GE-3} \\
\left( \omega +\varepsilon _{d\bar{\sigma}}-\Gamma _{2}\left( \omega \right)
\right) \langle \langle d_{\bar{\sigma}}^{\dagger };d_{\bar{\sigma}}\rangle
\rangle -\Delta _{2}\left( \omega \right) \langle \langle d_{\sigma };d_{%
\bar{\sigma}}\rangle \rangle  &=&1-U\langle \langle d_{\bar{\sigma}%
}^{\dagger }n_{d\sigma };d_{\bar{\sigma}}\rangle \rangle ,  \label{GE-4}
\end{eqnarray}%
where the following notations are introduced for clarity, $\Gamma _{1}\left( \omega \right)
=\sum_{k\beta }\frac{\left( \omega +\varepsilon _{-k\beta }\right) V_{k\beta
}^{\ast }V_{k\beta }}{E_{k\beta }^{2}},$ $\Gamma _{2}\left( \omega \right)
=\sum_{k\beta }\frac{\left( \omega -\varepsilon _{k\beta }\right) V_{-k\beta
}^{\ast }V_{k\beta }}{E_{k\beta }^{2}},$ $\Delta _{1}\left( \omega \right)
=\sum_{k\beta }\frac{\sigma \Delta \delta _{\beta ,S}V_{k\beta }^{\ast
}V_{-k\beta }^{\ast }}{E_{k\beta }^{2}}$ and $\Delta _{2}\left( \omega
\right) =\sum_{k\beta }\frac{\bar{\sigma}\Delta ^{\ast }\delta _{\beta
,S}V_{k\beta }V_{k\beta }}{E_{k\beta }^{2}}$ with $E_{k\beta }^{2}=\left(
\omega -\varepsilon _{k\beta }\right) \left( \omega +\varepsilon _{-k\beta
}\right) -\left\vert \Delta \right\vert ^{2}\delta _{\beta ,S}.$ To
proceed, it is convenient to transform the summation of $k$ into the
integral. For simplicity, in the following it is assumed that
the hybrid function is real and $k$-independent ($V_{k\beta}=V_{\beta}=V^{\ast}_{\beta}$), and the
superconducting gap is a real constant ($\Delta=\Delta^{\ast}$). Then, one obtains $\Gamma _{1}(\omega )=\Gamma _{2}(\omega
)=-i\Gamma _{N}-i\Gamma _{S}\rho _{S}\left( \omega \right) =-i\Gamma (\omega
)$ and the notations $\Delta _{1}(\omega )=-i\Gamma _{S}\rho _{S}\left(
\omega \right) \frac{\sigma \Delta }{\omega },\ \Delta _{2}(\omega
)=-i\Gamma _{S}\rho _{S}\left( \omega \right) \frac{\bar{\sigma}\Delta }{%
\omega }=-\Delta _{1}(\omega ),$ where $\Gamma _{N/S}=\pi |V_{N/S}|^{2}\rho
_{0}$ ($\rho _{0}=1/2D,$ $D$ is the half-bandwidth) and $\rho _{S}\left(
\omega \right) =\frac{|\omega| \theta \left( |\omega |-\Delta \right) }{\sqrt{%
\omega ^{2}-\Delta ^{2}}}+\frac{\omega \theta \left( \Delta -|\omega
|\right) }{i\sqrt{\Delta ^{2}-\omega ^{2}}}.$

Introducing the matrix notation
\begin{eqnarray}
\hat{G}_{d\sigma }\left( \omega \right)
=\left(
\begin{array}{cc}
\langle \langle d_{\sigma };d_{\sigma }^{\dagger }\rangle \rangle  &
\langle \langle d_{\sigma };d_{\bar{\sigma}}\rangle \rangle
\\
\langle\langle d_{\bar{\sigma}}^{\dagger };d_{\sigma }^{\dagger }\rangle \rangle&
\langle\langle d_{\bar{\sigma}}^{\dagger };d_{\bar{\sigma}}\rangle \rangle
\end{array}%
\right)  \label{GMD}
\end{eqnarray}
and
\begin{eqnarray}
\hat{F}_{d\sigma }\left( \omega \right) =\left(
\begin{array}{cc}
\langle \langle d_{\sigma }n_{d\bar{\sigma}};d_{\sigma }^{\dagger }\rangle
\rangle  & \langle \langle d_{\sigma }n_{d\bar{\sigma}};d_{\bar{\sigma}%
}\rangle \rangle  \\
-\langle \langle d_{\bar{\sigma}}^{\dagger }n_{d\sigma };d_{\sigma
}^{\dagger }\rangle \rangle  & -\langle \langle d_{\bar{\sigma}}^{\dagger
}n_{d\sigma };d_{\bar{\sigma}}\rangle \rangle
\end{array}
\right),  \label{FMD}
\end{eqnarray}
the Eqs.(\ref{GE-1})-(\ref{GE-4}) can be rewritten as a matrix form
\begin{equation}
\lbrack \hat{G}_{d\sigma }^{0}\left( \omega \right) ]^{-1}\hat{G}_{d\sigma
}\left( \omega \right) =I+U\hat{F}_{d\sigma }\left( \omega \right) ,
\label{APP-GM}
\end{equation}%
where $[\hat{G}_{d\sigma }^{0}\left( \omega \right) ]^{-1}=\left(
\begin{array}{cc}
\omega -\varepsilon _{d\sigma }-\hat{\Sigma}_{11}^{0}\left( \omega \right)
& -\hat{\Sigma}_{12}^{0}\left( \omega \right)  \\
-\hat{\Sigma}_{21}^{0}\left( \omega \right)  & \omega +\varepsilon _{d\bar{%
\sigma}}-\hat{\Sigma}_{22}^{0}\left( \omega \right)
\end{array}%
\right) $ and $\hat{\Sigma}_{11}^{0}\left( \omega \right) =\hat{\Sigma}%
_{22}^{0}\left( \omega \right) =-i\Gamma (\omega ),$ $\hat{\Sigma}%
_{12}^{0}\left( \omega \right) =\hat{\Sigma}_{21}^{0}\left( \omega \right)
=i\Gamma _{S}\rho _{S}\left( \omega \right) \frac{\sigma \Delta }{\omega }.$
This reproduces the formulas given in Refs.[\onlinecite{Bauer2007}] and [\onlinecite
{Bulla1998}].

It is well-known that Eq.(\ref{APP-GM}) is not closed since $\hat{F}_{d\sigma }\left( \omega \right) $ includes some high-order GFs. To proceed, one needs to employ the truncation approximation. In the following we present some brief steps in deriving the GFs in the HF and the Lacroix's approximate levels.

It is concise to treat $\hat{F}_{d\sigma }\left( \omega \right) $ by the cluster
expansion, \cite{Luo1999} e.g.,
\begin{equation}
\langle \langle d_{\sigma }n_{d\bar{\sigma}%
};d_{\sigma }^{\dagger }\rangle \rangle =\left\langle n_{d\bar{\sigma}%
}\right\rangle \langle \langle d_{\sigma };d_{\sigma }^{\dagger }\rangle
\rangle + \langle d_{\bar{\sigma}}d_{\sigma}\rangle \langle\langle d_{\bar{\sigma} }^{\dagger};
d_{\sigma}^{\dagger} \rangle\rangle+\langle \langle d_{\sigma }n_{d\bar{\sigma}};d_{\sigma }^{\dagger}\rangle \rangle_{c}.
\label{CE}
\end{equation}%
One can obtain
\begin{equation}
\hat{F}_{d\sigma }\left( \omega \right) = \hat{\Sigma}_{d\sigma }^{HF}\hat{G}%
_{d\sigma }\left( \omega \right) +\hat{F}_{d\sigma}\left( \omega \right)_c,
\label{FG}
\end{equation}%
where $\hat{\Sigma}_{d\sigma }^{HF}=\left(
\begin{array}{cc}
\langle n_{d\bar{\sigma}}\rangle  & \langle d_{\bar{\sigma}}d_{\sigma
}\rangle  \\
\langle d_{\sigma }^{\dagger }d_{\bar{\sigma}}^{\dagger }\rangle  & -\langle
n_{d\sigma }\rangle
\end{array}%
\right) $ represents the non-connected part, while $\hat{F}_{d\sigma}\left( \omega \right)_c =\left(
\begin{array}{cc}
\langle \langle d_{\sigma }n_{d\bar{\sigma}};d_{\sigma }^{\dagger }\rangle
\rangle _{c} & \langle \langle d_{\sigma }n_{d\bar{\sigma}};d_{\bar{\sigma}%
}\rangle \rangle _{c} \\
-\langle \langle d_{\bar{\sigma}}^{\dagger }n_{d\sigma };d_{\sigma
}^{\dagger }\rangle \rangle _{c} & -\langle \langle d_{\bar{\sigma}%
}^{\dagger }n_{d\sigma };d_{\bar{\sigma}}\rangle \rangle _{c}%
\end{array}%
\right) $ denotes connected (residual) part.
In Eq.(\ref{APP-GM}), $U\hat{F}%
_{d\sigma }\left( \omega \right) $ corresponds to the HF self-energy if $%
\hat{F}_{d\sigma}\left( \omega \right)_c $ is neglected,\cite{Vecino2003,Cuevas2001,Benjamin2007} and one
has
\begin{equation}
\left( \lbrack \hat{G}_{d\sigma }^{0}\left( \omega \right) ]^{-1}-U\hat{%
\Sigma}_{d\sigma }^{HF}\right) \hat{G}_{d\sigma }\left( \omega \right) =I.
\label{GE}
\end{equation}

Beyond the HF approximation, one needs to take $\hat{F}_{d\sigma}\left( \omega \right)_c $ into account, which is a bit complex but straightforward. Here our main aim is to
study the interplay between the superconductivity and the Kondo effect.
While superconductivity is already available in the superconducting
lead, the diagonal components in $\hat{F}_{d\sigma}\left( \omega \right)_c$ is sufficient to capture the Kondo effect in the Lacroix's approximation level.
Therefore, in the following we neglect anomalous high-order GFs like $\langle \langle d_{\sigma }n_{d\bar{\sigma}};d_{\bar{\sigma}}\rangle
\rangle _{c}$ and $\langle \langle d_{\bar{\sigma}}^{\dagger }n_{d\sigma
};d_{\sigma }^{\dagger }\rangle \rangle _{c}$ for simplification.
Immediately, one can obtain
\begin{equation}
\langle \langle d_{\bar{\sigma}}^{\dagger };d_{\sigma }^{\dagger }\rangle
\rangle =\frac{\hat{\Sigma}_{21}^{0}\left( \omega \right) +U\langle
d_{\sigma }^{\dagger }d_{\bar{\sigma}}^{\dagger }\rangle }{\omega
+\varepsilon _{d\bar{\sigma}}-\hat{\Sigma}_{22}^{0}\left( \omega \right)
+U\langle n_{d\sigma }\rangle }\langle \langle d_{\sigma };d_{\sigma
}^{\dagger }\rangle \rangle ,  \label{AGF-GF}
\end{equation}%
and
\begin{equation}
\left( \omega -\varepsilon _{d\sigma }-\hat{\Sigma}_{11}^{0}\left( \omega
\right) -U\langle n_{d\bar{\sigma}}\rangle \right) \langle \langle d_{\sigma
};d_{\sigma }^{\dagger }\rangle \rangle -\left( \hat{\Sigma}_{12}^{0}\left(
\omega \right) +U\langle d_{\bar{\sigma}}d_{\sigma }\rangle \right) \langle
\langle d_{\bar{\sigma}}^{\dagger };d_{\sigma }^{\dagger }\rangle \rangle
=1+\langle \langle d_{\sigma }n_{d\bar{\sigma}};d_{\sigma }^{\dagger
}\rangle \rangle _{c}.  \label{GF-AGF}
\end{equation}%
Above equations can be calculated self-consistently if we obtain $\langle
\langle d_{\sigma }n_{d\bar{\sigma}};d_{\sigma }^{\dagger }\rangle \rangle
_{c}$. For $\langle \langle d_{\sigma };d_{\bar{\sigma}}\rangle \rangle $ and $\langle \langle d_{\bar{%
\sigma}}^{\dagger };d_{\bar{\sigma}}\rangle \rangle $ one has a similar treatment (not shown here) for clarity.

To obtain the equation of motion of $\langle \langle d_{\sigma }n_{d\bar{\sigma}};d_{\sigma
}^{\dagger }\rangle \rangle _{c}$, we start from the equation of motion of
the Green's function $\langle \langle d_{\sigma }n_{d\bar{\sigma}};d_{\sigma
}^{\dagger }\rangle \rangle ,$
\begin{equation}
(\omega -\varepsilon _{d\sigma }-U)\langle \langle d_{\sigma }n_{d\bar{\sigma%
}};d_{\sigma }^{\dagger }\rangle \rangle =\langle n_{d\bar{\sigma}}\rangle
+\sum_{k\beta }V_{\beta }\langle \langle c_{k\sigma \beta }n_{d\bar{\sigma}%
};d_{\sigma }^{\dagger }\rangle \rangle +\sum_{k\beta }V_{\beta }\left(
\langle \langle d_{\bar{\sigma}}^{\dagger }c_{-k\bar{\sigma}\beta }d_{\sigma
};d_{\sigma }^{\dagger }\rangle \rangle -\langle \langle c_{-k\bar{\sigma}%
\beta }^{\dagger }d_{\bar{\sigma}}d_{\sigma };d_{\sigma }^{\dagger }\rangle
\rangle \right) ,  \label{GF-dnd}
\end{equation}%
which involves other high-order GFs $\langle \langle c_{k\sigma \beta }n_{d%
\bar{\sigma}};d_{\sigma }^{\dagger }\rangle \rangle ,$ $\langle \langle d_{%
\bar{\sigma}}^{\dagger }c_{-k\bar{\sigma}\beta }d_{\sigma };d_{\sigma
}^{\dagger }\rangle \rangle ,$ and $\langle \langle c_{-k\bar{\sigma}\beta
}^{\dagger }d_{\bar{\sigma}}d_{\sigma };d_{\sigma }^{\dagger }\rangle
\rangle $, whose equations of motion are given as follows
\begin{eqnarray}
(\omega -\varepsilon _{k\beta })\langle \langle c_{k\sigma \beta }n_{d\bar{%
\sigma}};d_{\sigma }^{\dagger }\rangle \rangle  &=&V_{\beta }\langle \langle
d_{\sigma }n_{d\bar{\sigma}};d_{\sigma }^{\dagger }\rangle \rangle +\sigma
\Delta \delta _{\beta ,S}\langle \langle c_{-k\bar{\sigma}\beta }^{\dagger
}n_{d\bar{\sigma}};d_{\sigma }^{\dagger }\rangle \rangle   \notag \\
&&+\sum_{k^{\prime }\beta ^{\prime }}V_{\beta ^{\prime }} \langle
\langle d_{\bar{\sigma}}^{\dagger }c_{-k^{\prime }\bar{\sigma}\beta ^{\prime
}}c_{k\sigma \beta };d_{\sigma }^{\dagger }\rangle \rangle -\sum_{k^{\prime
}\beta ^{\prime }}V_{\beta ^{\prime }}\langle \langle c_{-k^{\prime }\bar{%
\sigma}\beta ^{\prime }}^{\dagger }d_{\bar{\sigma}}c_{k\sigma \beta
};d_{\sigma }^{\dagger }\rangle \rangle,  \label{GF-cnd}
\end{eqnarray}%
\begin{eqnarray}
(\omega -\omega _{1,k\sigma \beta })\langle \langle d_{\bar{\sigma}%
}^{\dagger }c_{-k\bar{\sigma}\beta }d_{\sigma };d_{\sigma }^{\dagger }\rangle
\rangle  &=&\langle d_{\bar{\sigma}}^{\dagger }c_{-k\bar{\sigma}\beta
}\rangle +V_{\beta }\langle \langle d_{\sigma }n_{\bar{\sigma}};d_{\sigma
}^{\dagger }\rangle \rangle -\sum_{k^{\prime }\beta ^{\prime }}V_{\beta
^{\prime }}\langle \langle c_{-k^{\prime }\bar{\sigma}\beta ^{\prime
}}^{\dagger }c_{-k\bar{\sigma}\beta }d_{\sigma };d_{\sigma }^{\dagger
}\rangle \rangle   \notag \\
&&+\sum_{k^{\prime }\beta ^{\prime }}V_{\beta ^{\prime }} \langle
\langle d_{\bar{\sigma}}^{\dagger }c_{-k\bar{\sigma}\beta }c_{k^{\prime
}\sigma \beta ^{\prime }};d_{\sigma }^{\dagger }\rangle \rangle +\bar{\sigma}%
\Delta \delta _{\beta ,S}\langle \langle d_{\bar{\sigma}}^{\dagger
}c_{k\sigma \beta }^{\dagger }d_{\sigma };d_{\sigma }^{\dagger }\rangle
\rangle,  \label{GF-dcdd}
\end{eqnarray}%
\begin{eqnarray}
(\omega -\omega _{2,k\sigma \beta })\langle \langle c_{-k\bar{\sigma}\beta
}^{\dagger }d_{\bar{\sigma}}d_{\sigma };d_{\sigma }^{\dagger }\rangle
\rangle  &=&\langle c_{-k\bar{\sigma}\beta }^{\dagger }d_{\bar{\sigma}%
}\rangle -V_{\beta } \langle \langle d_{\sigma }n_{\bar{\sigma}%
};d_{\sigma }^{\dagger }\rangle \rangle -\bar{\sigma}\Delta\delta
_{\beta ,S}\langle \langle c_{k\sigma \beta }d_{\bar{\sigma}}d_{\sigma
};d_{\sigma }^{\dagger }\rangle \rangle   \notag \\
&&+\sum_{k^{\prime }\beta ^{\prime }}V_{\beta ^{\prime }} \langle
\langle c_{-k\bar{\sigma}\beta }^{\dagger }c_{-k^{\prime }\bar{\sigma}\beta
^{\prime }}d_{\sigma };d_{\sigma }^{\dagger }\rangle \rangle
+\sum_{k^{\prime }\beta ^{\prime }}V_{\beta ^{\prime }}\langle
\langle c_{-k\bar{\sigma}\beta }^{\dagger }d_{\bar{\sigma}}c_{k^{\prime
}\sigma \beta ^{\prime }};d_{\sigma }^{\dagger }\rangle \rangle,
\label{GF-cddd}
\end{eqnarray}%
where $\omega _{1,k\sigma \beta }=\varepsilon _{-k\beta }+\varepsilon
_{d\sigma }-\varepsilon _{d\bar{\sigma}}$ and $\omega _{2,k\sigma \beta
}=-\varepsilon _{-k\beta }+\varepsilon _{d\bar{\sigma}}+\varepsilon _{d\sigma
}+U.$
In order to obtain $\langle \langle d_{\sigma }n_{d\bar{\sigma}};d_{\sigma
}^{\dagger }\rangle \rangle _{c}$, the other high-order GFs involved in
Eqs.(\ref{GF-cnd})-(\ref{GF-cddd}) have been treated by the cluster expansion,
e.g., $\langle \langle c_{k\sigma\beta}n_{d\bar{\sigma}};d_{\sigma }^{\dagger
}\rangle \rangle \approx \langle n_{d\bar{\sigma}}\rangle \langle
\langle c_{k\sigma\beta };d_{\sigma }^{\dagger }\rangle \rangle +\langle \langle
c_{k\sigma\beta }n_{d\bar{\sigma}};d_{\sigma }^{\dagger }\rangle \rangle _{c}.$
Here, the additional approximation taken is to neglect all
superconducting correlation functions involved QD-leads pairing. For the same reason, the
possible pairing corrections brought from the higher order GFs like
$\langle \langle c_{-k\bar{\sigma}\beta }^{\dagger
}n_{d\bar{\sigma}};d_{\sigma }^{\dagger }\rangle \rangle $,
$\langle \langle d_{\bar{\sigma}}^{\dagger
}c_{-k\sigma \beta }^{\dagger }d_{\sigma };d_{\sigma }^{\dagger }\rangle
\rangle$ and $\langle \langle c_{-k\sigma \beta }d_{\bar{\sigma}}d_{\sigma
};d_{\sigma }^{\dagger }\rangle \rangle $ are also neglected.

After a long but straightforward calculation, one has%
\begin{eqnarray}
&&\left( \omega -\varepsilon _{d\sigma }-U\left( 1-\left\langle n_{d\bar{%
\sigma}}\right\rangle \right) \right) \langle \langle d_{\sigma }n_{d\bar{%
\sigma}};d_{\sigma }^{\dagger }\rangle \rangle _{c}=U[\left\langle n_{d\bar{%
\sigma}}\right\rangle \left( 1-\left\langle n_{d\bar{\sigma}}\right\rangle
\right)- \langle d_{\bar{\sigma}%
}d_{\sigma}\rangle \langle d_{\sigma}^{\dag}d_{\bar{\sigma}%
}^{\dagger}\rangle ] \langle \langle d_{\sigma };d_{\sigma }^{\dagger }\rangle \rangle
\notag \\
&&+\left[  \varepsilon_{d\sigma}+\varepsilon_{d\bar{\sigma}}+U\left(
1-\left\langle n_{d\bar{\sigma}}\right\rangle +\left\langle n_{d\sigma
}\right\rangle \right)  \right]  \left\langle d_{\bar{\sigma}}d_{\sigma
}\right\rangle \langle\langle d_{\bar{\sigma}}^{\dagger};d_{\sigma}^{\dag
}\rangle\rangle+\sum_{k\beta }V_{\beta }\langle \langle c_{k\sigma \beta }n_{d
\bar{\sigma}};d_{\sigma }^{\dagger }\rangle \rangle _{c}
\notag \\
&&-\sum_{k\beta
}V_{\beta }\langle \langle c_{-k\bar{\sigma}\beta }^{\dagger }d_{\bar{\sigma}
}d_{\sigma };d_{\sigma }^{\dagger }\rangle \rangle _{c}+\sum_{k\beta
}V_{\beta }\langle \langle d_{\bar{\sigma}}^{\dagger }c_{-k\bar{\sigma
}\beta }d_{\sigma };d_{\sigma }^{\dagger }\rangle \rangle _{c},
\label{GF-dndc}
\end{eqnarray}
Likewise, the more higher order connected GFs read
\begin{eqnarray}
(\omega -\varepsilon _{k\beta })\langle \langle c_{k\sigma \beta }n_{d\bar{%
\sigma}};d_{\sigma }^{\dagger }\rangle \rangle _{c} &\approx &V_{\beta
}\langle \langle d_{\sigma }n_{d\bar{\sigma}};d_{\sigma }^{\dagger }\rangle
\rangle _{c},  \label{CGF-cnd} \\
(\omega -\omega _{1,k\sigma \beta })\langle \langle d_{\bar{\sigma}%
}^{\dagger }c_{-k\bar{\sigma}\beta }d_{\sigma };d_{\sigma }^{\dagger }\rangle
\rangle _{c} &\approx &A_{1,k\sigma \beta }\langle \langle d_{\sigma
};d_{\sigma }^{\dagger }\rangle \rangle +B_{1,k\sigma \beta }\langle \langle
d_{\sigma }n_{d\bar{\sigma}};d_{\sigma }^{\dagger }\rangle \rangle _{c},
\label{CGF-dcd} \\
(\omega -\omega _{2,k\sigma \beta })\langle \langle c_{-k\bar{\sigma}\beta
}^{\dagger }d_{\bar{\sigma}}d_{\sigma };d_{\sigma }^{\dagger }\rangle
\rangle _{c} &\approx &A_{2,k\sigma \beta }\langle \langle d_{\sigma
};d_{\sigma }^{\dagger }\rangle \rangle -B_{2,k\sigma \beta }\langle \langle
d_{\sigma }n_{d\bar{\sigma}};d_{\sigma }^{\dagger }\rangle \rangle _{c},
\label{CGF-cdd}
\end{eqnarray}%
where%
\begin{eqnarray}
A_{1,k\sigma \beta } &=&\left( \omega _{1,k\sigma \beta }-\varepsilon
_{d\sigma }-U\left\langle n_{d\bar{\sigma}}\right\rangle \right) \langle d_{%
\bar{\sigma}}^{\dagger }c_{-k\bar{\sigma}\beta }\rangle +V_{\beta
}\left\langle n_{d\bar{\sigma}}\right\rangle -\sum_{k^{\prime }\beta
^{\prime }}V_{\beta ^{\prime }}\langle c_{-k^{\prime }\bar{\sigma}\beta
^{\prime }}^{\dagger }c_{-k\bar{\sigma}\beta }\rangle ,  \label{ca1} \\
A_{2,k\sigma \beta } &=&\left( \omega _{2,k\sigma \beta }-\varepsilon
_{d\sigma }-U\left\langle n_{d\bar{\sigma}}\right\rangle \right) \langle c_{-k%
\bar{\sigma}\beta }^{\dagger }d_{\bar{\sigma}}\rangle -V_{\beta }
\left\langle n_{d\bar{\sigma}}\right\rangle +\sum_{k^{\prime }\beta
^{\prime }}V_{\beta ^{\prime }}\langle c_{-k\bar{\sigma}\beta
}^{\dagger }c_{-k^{\prime }\bar{\sigma}\beta ^{\prime }}\rangle ,  \label{ca2}
\end{eqnarray}%
and $B_{1,k\sigma \beta }=V_{\beta }-U\langle d_{\bar{\sigma}}^{\dagger }c_{-k%
\bar{\sigma}\beta }\rangle ,$ $B_{2,k\sigma \beta }=V_{\beta }
+U\langle c_{-k\bar{\sigma}\beta }^{\dagger }d_{\bar{\sigma}}\rangle $. The
trunction scheme taken above in the derivation of Eqs. (\ref{GF-dndc})-(\ref%
{CGF-cdd}) is essentially the famous Lacroix's approximation, which is
sufficient to capture qualitatively the Kondo effect. \cite{Luo1999,Lacroix1981}

By substituting Eqs.(\ref{CGF-cnd})-(\ref{CGF-cdd}) into Eq.(\ref{GF-dndc}),
then going back to Eq.(\ref{GF-AGF}), in combination with Eq.(\ref{AGF-GF}),
one has
\begin{equation}
\langle \langle d_{\sigma };d_{\sigma }^{\dagger }\rangle \rangle =\frac{1}{%
R_{\sigma }(\omega )-U[Q_{\sigma }(\omega )+T_{\sigma }(\omega )]/P_{\sigma }(\omega )},  \label{GFS}
\end{equation}
where
\begin{eqnarray}
R_{\sigma }(\omega ) &=&\omega -\varepsilon _{d\sigma }+i\Gamma (\omega
)-U\langle n_{d\bar{\sigma}}\rangle -\frac{( \hat{\Sigma}%
_{21}^{0}\left( \omega \right) +U\langle d_{\bar{\sigma}}d_{\sigma }\rangle
) ( \hat{\Sigma}_{21}^{0}\left( \omega \right) +U\langle
d_{\sigma }^{\dagger }d_{\bar{\sigma}}^{\dagger }\rangle) }{\omega
+\varepsilon _{d\bar{\sigma}}+i\Gamma (\omega )+U\langle n_{d\sigma }\rangle
},  \label{Qsigma} \\
P_{\sigma }(\omega ) &=&\omega -\varepsilon _{d\sigma }-U\left( 1-\langle
n_{d\bar{\sigma}}\rangle \right) +3i\Gamma (\omega )+U(A_{1\sigma}(\omega)-A_{2\sigma}(\omega)),  \label{Psigma} \\
Q_{\sigma }(\omega ) &=&\left( \omega -\varepsilon _{d\sigma }-U\langle n_{d\bar{\sigma}}\rangle
\right) (A_{1\sigma}(\omega)-A_{2\sigma}(\omega))+U[\langle n_{d\bar{\sigma}}\rangle ( 1-\langle
n_{d\bar{\sigma}}\rangle)-\langle d_{\bar{\sigma}%
}d_{\sigma}\rangle \langle d_{\sigma}^{\dag}d_{\bar{\sigma}%
}^{\dagger}\rangle] \notag \\
&&-2i\Gamma (\omega )\langle n_{d\bar{\sigma}%
}\rangle - (B_{1\sigma}(\omega)+B_{2\sigma}(\omega)),  \label{Rsigma} \\
T_{\sigma}\left(  \omega\right)&=&\left[  \varepsilon_{d\sigma
}+\varepsilon_{d\bar{\sigma}}+U\left(  1+\left\langle n_{d\sigma}\right\rangle
-\left\langle n_{d\bar{\sigma}}\right\rangle \right)  \right]  \frac{ \left\langle
d_{\bar{\sigma}}d_{\sigma}\right\rangle(\hat{\Sigma}_{21}^{0}\left( \omega \right)
+U\langle d_{\sigma}^{\dagger}d_{\bar{\sigma}}^{\dag}\rangle)}{\omega
+\varepsilon _{d\bar{\sigma}}+i\Gamma (\omega )+U\left\langle n_{d\sigma
}\right\rangle }.\label{ccc}
\end{eqnarray}%
In $P_{\sigma }(\omega )$ and $Q_{\sigma }(\omega )$, the notation
\begin{eqnarray}
A_{i\sigma }( \omega )  &=&\sum_{k\beta }\frac{%
V_{\beta }\langle d_{\bar{\sigma}}^{\dagger }c_{-k\bar{\sigma}\beta }\rangle
}{z_{+}-\omega _{i,k\sigma \beta }}  \notag \\
&=&\frac{i}{2\pi }\sum_{k\beta }\frac{1}{z_{+}-\omega
_{i,k\sigma \beta }}\int d\omega ^{\prime }f( \omega ^{\prime })
V_{\beta }(\langle \langle c_{-k\bar{\sigma}\beta };d_{\bar{\sigma}}^{\dagger
}\rangle \rangle _{\omega ^{\prime }}^{r}-\langle \langle c_{-k\bar{\sigma}%
\beta };d_{\bar{\sigma}}^{\dagger }\rangle \rangle _{\omega ^{\prime }}^{a})
\notag \\
&=&\frac{i}{2\pi ^{2}}\sum_{\beta  }\int \int d\omega
^{\prime }d\varepsilon \Gamma _{\beta }f( \omega ^{\prime })
\frac{\frac{( z_{+}^{\prime }+\varepsilon ) \langle \langle d_{%
\bar{\sigma}};d_{\bar{\sigma}}^{\dagger }\rangle \rangle _{\omega ^{\prime
}}^{r}-\delta _{\beta ,S}\bar{\sigma}\Delta \langle \langle d_{\sigma
}^{\dagger };d_{\bar{\sigma}}^{\dagger }\rangle \rangle _{\omega ^{\prime
}}^{r}}{( z_{+}^{\prime }-\varepsilon ) ( z_{+}^{\prime
}+\varepsilon ) -\delta _{\beta ,S}\Delta ^{2}}-\frac{(
z_{-}^{\prime }+\varepsilon ) \langle \langle d_{\bar{\sigma}};d_{\bar{%
\sigma}}^{\dagger }\rangle \rangle _{\omega ^{\prime }}^{a}-\delta _{\beta
,S}\bar{\sigma}\Delta \langle \langle d_{\sigma }^{\dagger };d_{\bar{\sigma}%
}^{\dagger }\rangle \rangle _{\omega ^{\prime }}^{a}}{( z_{-}^{\prime
}-\varepsilon ) ( z_{-}^{\prime }+\varepsilon ) -\delta
_{\beta ,S}\Delta ^{2}}}{z_{+}-\varepsilon _{i\sigma}}
\notag \\
\label{EVdc}
\end{eqnarray}%
where $z_{\pm }=\omega \pm i0^{+}$, $\varepsilon _{1\sigma }=\varepsilon
+\varepsilon _{d\sigma }-\varepsilon _{d\bar{\sigma}}$, and $\varepsilon
_{2\sigma }=-\varepsilon +\varepsilon _{d\bar{\sigma}}+\varepsilon _{d\sigma
}+U.$ The average value $\langle d_{\bar{\sigma }}^{\dagger
}c_{-k\bar{\sigma }\beta }\rangle$ is calculated by the
spectral theorem $\langle d_{\bar{\sigma}}^{\dagger }c_{-k\bar{\sigma%
}\beta }\rangle =\frac{i}{2\pi }\int d\omega f(\omega )(\langle \langle c_{-k%
\bar{\sigma}\beta };d_{\bar{\sigma}}^{\dagger }\rangle \rangle _{\omega
}^{r}-\langle \langle c_{-k\bar{\sigma}\beta };d_{\bar{\sigma}}^{\dagger
}\rangle \rangle _{\omega }^{a})$, and $\langle c_{-k\bar{\sigma }%
\beta }^{\dagger }d_{\bar{\sigma}}\rangle =\langle d_{\bar{\sigma }}^{\dagger
}c_{-k\bar{\sigma }\beta }\rangle$ is taken in Eq.(\ref{Psigma}) and Eq.(\ref{Rsigma}).
The GF obtained is $\langle \langle c_{k\bar{\sigma}%
\beta };d_{\bar{\sigma}}^{\dagger }\rangle \rangle _{\omega }^{r(a)}=\frac{%
(z_{\pm }+\varepsilon _{-k\beta })V_{\beta }\langle \langle d_{\bar{\sigma}%
};d_{\bar{\sigma}}^{\dagger }\rangle \rangle _{\omega }^{r(a)}-\delta
_{\beta ,S}\bar{\sigma}\Delta V_{\beta }\langle \langle d_{\sigma }^{\dagger
};d_{\bar{\sigma}}^{\dagger }\rangle \rangle _{\omega }^{r(a)}}{(z_{\pm
}-\varepsilon _{k\beta })(z_{\pm }+\varepsilon _{-k\beta })-\delta _{\beta
,S}\Delta ^{2}}.$ Similarly, we can obtain
\begin{eqnarray}
B_{i\sigma }( \omega )  &=&\sum_{kk^{\prime }\beta \beta ^{\prime
}}\frac{|V_{\beta }|^{2}\langle c_{-k^{\prime }\bar{\sigma}\beta
^{\prime }}^{\dagger }c_{-k\bar{\sigma}\beta }\rangle }{z_{+}-\omega
_{i,k\sigma \beta }}  \notag \\
&=&\frac{i}{2\pi }\sum_{kk^{\prime }\beta \beta ^{\prime }}\int
d\omega ^{\prime }f(\omega ^{\prime })\frac{V_{\beta }V_{\beta ^{\prime
}}(\langle \langle c_{-k\bar{\sigma}\beta };c_{-k^{\prime }\bar{\sigma}\beta
^{\prime }}^{\dagger }\rangle \rangle _{\omega ^{\prime }}^{r}-\langle
\langle c_{-k\bar{\sigma}\beta };c_{-k^{\prime }\bar{\sigma}\beta ^{\prime
}}^{\dagger }\rangle \rangle _{\omega ^{\prime }}^{a})}{z_{+}-\omega
_{i,k\sigma \beta }}  \notag \\
&\approx &\frac{1}{\pi }\int d\varepsilon \frac{\Gamma _{N}f(\varepsilon
)( 1-i\Gamma _{N}\langle \langle d_{\bar{\sigma}};d_{\bar{\sigma}%
}^{\dagger }\rangle \rangle _{\varepsilon }^{r}) }{z_{+}-\varepsilon
_{i\sigma }}+\frac{i}{2\pi ^{2}}\int \int d\omega^{\prime}d\varepsilon 
f(\omega ^{\prime })\Gamma _{S}(\varepsilon )\frac{\frac{z_{+}^{\prime
}+\varepsilon }{(z_{+}^{\prime }-\varepsilon )(z_{+}^{\prime }+\varepsilon
)-\Delta ^{2}}-\frac{z_{-}^{\prime }+\varepsilon }{(z_{-}^{\prime
}-\varepsilon )(z_{-}^{\prime }+\varepsilon )-\Delta ^{2}}}{z_{+}-\varepsilon
_{i\sigma}},
\notag \\
\label{EVCCD}
\end{eqnarray}%
where $\langle c_{-k^{\prime }\bar{\sigma}\beta ^{\prime }}^{\dagger }c_{-k%
\bar{\sigma}\beta }\rangle =\frac{i}{2\pi }\int d\omega f(\omega )(\langle
\langle c_{-k\bar{\sigma}\beta };c_{-k^{\prime }\bar{\sigma}\beta ^{\prime
}}^{\dagger }\rangle \rangle _{\omega }^{r}-\langle \langle c_{-k\bar{\sigma}%
\beta };c_{-k^{\prime }\bar{\sigma}\beta ^{\prime }}^{\dagger }\rangle
\rangle _{\omega }^{a})$, and the GFs involved can be approximately obtained
with $\sum_{\beta \beta ^{\prime }(=N,S)}\langle \langle c_{k\bar{\sigma}%
\beta };c_{k^{\prime }\bar{\sigma}\beta ^{\prime }}^{\dagger }\rangle
\rangle _{\omega }^{r(a)}\approx \frac{\delta _{kk^{\prime }}}{z_{\pm
}^{\prime }-\varepsilon _{kN}}+\frac{|V_{N}|^{2}\langle \langle d_{\bar{%
\sigma}};d_{\bar{\sigma}}^{\dagger }\rangle \rangle _{\omega }^{r(a)}}{%
(z_{\pm }^{\prime }-\varepsilon _{kN})(z_{\pm }^{\prime }-\varepsilon
_{k^{\prime }N})}+\frac{(z_{\pm }^{\prime }+\varepsilon _{kS})\delta
_{kk^{\prime }}}{(z_{\pm }^{\prime }-\varepsilon _{kS})(z_{\pm }^{\prime
}+\varepsilon _{-kS})-\Delta ^{2}}.$
\end{widetext}

\end{document}